\documentclass[10pt,a4paper]{article}
\usepackage{fullpage}

\usepackage[utf8]{inputenc}

\usepackage{amsmath,amsfonts,bm,amssymb,amsthm,dsfont}
\usepackage[round,sort&compress]{natbib}
\usepackage{graphicx}
\usepackage{xcolor}
\usepackage[textsize=footnotesize, backgroundcolor=white,bordercolor=red,linecolor=black]{todonotes}
\usepackage{hyperref}
\usepackage{wrapfig}
\usepackage{booktabs}
\usepackage{multirow}
\usepackage{xspace}
\usepackage{rotating}
\usepackage{pifont}
\usepackage{authblk}

\usepackage[T1]{fontenc}

\renewcommand{\vec}[1]{\mathbf{#1}}

\newcommand{\expect}[1]{\ensuremath{\mathds{E}[#1]}}
\DeclareMathOperator*{\argmax}{arg\,max}

\DeclareMathAlphabet{\mathcal}{OMS}{cmsy}{m}{n}

\newcommand{\model}{\mathcal{M}}

\newcommand{\BF}{\text{BF}}

\newcommand{\method}{BNDD\xspace}

\newcommand{\dnorm}[2]{\mathcal{N}\left(#1, #2 \right)}
\newcommand{\dgp}[2]{\mathcal{GP}\left(#1, #2 \right)}

\newcommand*{\email}[1]{%
    \normalsize\href{mailto:#1}{#1}\par
    }

\title{Bayesian nonparametric discontinuity design} 
\author{Max Hinne, David Leeftink, Marcel A. J. van Gerven, and Luca Ambrogioni}
\affil{Radboud University, Nijmegen, The Netherlands \\ \email{m.hinne@donders.ru.nl}
}

\begin{document}

\maketitle

\begin{abstract} 
    Quasi-experimental research designs, such as regression discontinuity and interrupted time series, allow for causal inference in the absence of a randomized controlled trial, at the cost of additional assumptions. In this paper, we provide a framework for discontinuity-based designs using Bayesian model comparison and Gaussian process regression, which we refer to as `Bayesian nonparametric discontinuity design', or BNDD for short. \method addresses the two major shortcomings in most implementations of such designs: overconfidence due to implicit conditioning on the alleged effect, and model misspecification due to reliance on overly simplistic regression models. With the appropriate Gaussian process covariance function, our approach can detect discontinuities of any order, and in spectral features. We demonstrate the usage of \method in simulations, and apply the framework to determine the effect of running for political positions on longevity, of the effect of an alleged historical phantom border in the Netherlands on Dutch voting behaviour, and of Kundalini Yoga meditation on heart rate.
\end{abstract}



\section{Introduction}
\label{sec:intro}

The bread and butter of scientific research is the randomized-controlled trial (RCT)~\citep{Hill1952}. In this design, the sample population is randomly divided into two groups; one that is manipulated (e.g. a drug is administered or a treatment is performed), while the other is left unchanged. RCT allows one to perform \emph{causal inference}, and learn about the causal effect of the intervention~\citep{Pearl2009,Imbens2015}. However, in practice there may be several insurmountable ethical or pragmatic hurdles that deter one from using RCT, such as ethical or pragmatic concerns. Luckily, all is not lost for experimental design. There exist several \emph{quasi}-experimental designs (QEDs) that replace random assignment with deterministic assignment, which still allow for causal inferences, but at the cost of additional assumptions~\citep{Shadish2002}. Prominent examples are regression discontinuity (RD) and interrupted time series (ITS) designs, that assign a sample to one of the two groups based on it passing a threshold on an assignment variable~\citep{Campbell1963,Lauritzen2001,McDowall1980}. The idea behind these approaches is that, around the assignment threshold, observations are distributed essentially randomly, so that locally the conditions of RCT are recreated~\citep{Lee2010,Imbens2008}. The methodological pipeline of quasi-experimental designs like these generally consists of three steps~\citep{Rischard2021}. First, a regression (typically linear) is fit to each of the two groups individually. Next, the regressions are extrapolated to the threshold (RD), or to the entire post-intervention range (ITS). Finally, the difference between the extrapolations of the two groups is taken as the effect size of the intervention. A straightforward statistical test can be applied to check whether the effect is present. 

Here, we provide a novel framework for such approaches, which we call `Bayesian nonparametric discontinuity design' (BNDD). The main innovations of \method are: First, we frame the problem of detecting an effect as Bayesian model comparison. Instead of comparing the pre- and post-intervention regressions, we introduce a continuous model and a discontinuous model. In the discontinuous model, observations before and after the intervention are assumed to be independent, while in the continuous model this assumption is lifted. We quantify the evidence in favor of either model, rather than only for the alternative model, via Bayesian model comparison~\citep{Wagenmakers2007}. This enables the computation of the marginal effect size, which provides a more nuanced estimate compared to implicitly conditioning on the alternative model~\citep{Hoeting1999}. Furthermore, the model comparison approach automatically penalizes the discontinuous model for its additional flexibility~\citep{MacKay2002}. Second, we use Gaussian process (GP) regression to avoid strong parametric assumptions. The result is a flexible model that can capture nonlinear interactions between the predictor and outcome variables. Traditional assumptions, such as linearity, can still be implemented in our model by using the appropriate covariance function. At the same time, much more expressive covariance functions can be used, such as the spectral mixture kernel~\citep{Wilson2013}, that better capture long-range correlations, and lead to more accurate inference. Lastly, in most discontinuity-based methods for quasi-experimental design, a bandwidth parameter determines the trade-off between estimation reliability and the local randomness assumptions that are needed to draw causal inferences~\citep{Geneletti2015}. In \method, all observations are used to estimate both the continuous and discontinuous model, but by optimizing the length-scale parameter of the GP covariance functions we control the sensitivity to different types of discontinuities and adherence to locality assumptions. 

\section{Related work}

\label{sec:relwork}

While quasi-experimental designs have been around since the 1960s~\citep{Thistlethwaite1960,Campbell1963}, recently there has been a renewed interest in this class of methods~\citep{Imbens2008,Choi2017}, in particular in epidemiology~\citep{Harris2004} and education~\citep{Bloom2012}. Researchers from different domains are promoting the use of QED~\citep{Marinescu2018,Moscoe2015,Liu2021}, which has prompted several extensions of classical QEDs. For instance, several authors have proposed to use Bayesian models for QED~\citep{Geneletti2015,Freni-Sterrantino2019}. By assuming a prior distribution for alleged effect size and using Bayes' theorem, these studies provide an explicit descriptions of the estimation uncertainty. In contrast to our work, these methods focus on the estimation of the treatment effect instead of model comparison, and typically assume restrictive parametric forms. Other studies have considered nonparametric alternatives to linear models. For example,~\citet{Hahn2001} use locally linear nonparametric regression. Alternatively, one can use kernel methods that compute a smoothly weighted average of the data points to create an interpolated regression that does not depend on a specific parametric form \citep{Bloom2012}. Most similar to \method are the approaches by~\citet{Rischard2021,Branson2019}. Here, instead of fitting a parametric form such as linear regression, the regression is modelled by a GP, which results in a flexible, nonparametric model and more accurate effect size estimates compared to when using linear regression. \method uses GP regression as well, but whereas~\citet{Rischard2021,Branson2019} focus on the inference of the magnitude of the treatment effect, we first determine whether an effect is present at all using Bayesian model comparison~\citep{Wagenmakers2007}, and we use Bayesian model averaging \citep{Hoeting1999} to reduce the overconfidence that follows from conditioning on the alternative model or particular covariance functions. Consequently, \method is less prone to false positives, is able to detect discontinuities in derivatives of the latent function rather than in the function per s\'{e}, and using the spectral mixture kernel our approach is well-suited for detecting changes in time series, which is crucial in ITS design.

\section{Discontinuity-based causal inference} 
\label{sec:bnpqed}

We provide a brief introduction of the background of causal inference using RD and ITS designs, for a more in-depth discussion, we refer to e.g.~\citet{Bernal2017,McDowall1980}. The detection of a causal effect is naturally formulated using the potential outcomes framework by~\citet{Rubin1974}, which assumes that for each individual in the study both the outcome of the treatment and its alternative can potentially be observed. 

Consider an observation $i$ (with or without temporal ordering) with independent variable $x_i\in \mathds{R}^p$ and response $y_i \in\mathds{R}^q$ (we will assume $p=q=1$, but multidimensional extensions are straightforward). In addition, we observe an indicator variable $z_i$, where $z_i=1$ denotes the intervention of interest has been applied to case $i$, and $z_i=0$ indicates it has not. The outcome depends on treatment, so
\begin{equation}
    y_i = \begin{cases}
    y_i(0) & \text{if $z_i=0$,}\\
    y_i(1) & \text{if $z_i=1$.}
    \end{cases}
\end{equation} The individual causal effect is defined as the difference between these two potential outcomes, that is $d_i = y_i(1) - y_i(0)$. Since we only ever observe one outcome, the individual causal effect is out of reach, so in RD design we focus on the \emph{average causal effect} (ACE) instead, defined by the differences in the expectations:
\begin{equation}
    d_{\text{ACE}} = \expect{y(1)} - \expect{y(0)} \enspace.
\end{equation} In the randomized controlled trial, the assignment of treatment $z_i$ is random, so that all differences other than due to the treatment are integrated out in these expectations~\citep{Bloom2012}. In QED designs such as RD and ITS however, the allocation to intervention or control group is based on a threshold $x_0$~\citep{OKeeffe2016}:
\begin{equation}
    z_i = \begin{cases}
    1 & \text{if $x_i \geq x_0$ and} \\
    0 & \text{otherwise.}
    \end{cases}
\end{equation} This changes how the ACE is computed, which for RD design becomes~\citep{Imbens2012,Lee2010}:
\begin{equation}\label{eq:rd_effectsize}
    \begin{split}
    d_{\text{RD}} &= \expect{y_i(1) - y_i(0) \mid x_i = x_0}\\
     &= \lim_{x\downarrow x_0} \expect{y_i | x_i = x_0} - \lim_{x\uparrow x_0} \expect{y_i | x_i = x_0} \enspace, 
    \end{split}
\end{equation} provided the distributions of $y_i$ given $x_i$ are continuous in $x$, and the conditional expectations $\expect{y_i(1)\mid x_i}$ and $\expect{y_i(0)\mid x_i}$ exist. 

For interrupted time series, there are no post-intervention control observations, as all post-threshold observations $x_i\geq x_0$ are in the intervention group. Here, the causal estimand becomes the \emph{average effect of the treatment on the treated} (ATT)~\citep{Kim2016}:
\begin{equation}\label{eq:its_effectsize}
    \begin{split}
    d_{\text{ITS}}(x_i)   &= \expect{y_i(1) - y_i(0) \mid x_i \geq x_0}\\
                        &= \expect{y_i\mid x_i, D} - \expect{y_i\mid x_i, D^0}\enspace,
    \end{split}
\end{equation} for $x_i \geq x_0$, $D=\{(x_i, y_i)\}_{i=1}^n$, and $D^0=\{(x_i, y_i)\}_{x_i < x_0}$. Intuitively, this measure of effect size is the difference between the extrapolation based on the pre-intervention data, and the actual post-intervention observations. Due the the reliance on extrapolation, it is crucial that correct assumptions are made on the functional form. For example, assuming linearity will lead to a biased ATT estimate if this does not describe the functional form well. 

Importantly, for both approaches we assume there are no confounding variables that affect the relationship between $x$ and $y$ (for a more in-depth discussion of RD design, see~\citet{Geneletti2015}).

\section{Bayesian nonparametric discontinuity design}

In standard RD and ITS analyses, causal conclusions are drawn by estimating the effect $d$ and testing whether this differs from zero. Instead, we perform Bayesian model comparison to see whether the data are better supported by the alternative model $\model_1$, that claims an effect is present, than by the null model $\model_0$, in which such an effect is absent. The result of the model comparison is quantified by the Bayes factor~\citep{Kass1995}:
\begin{equation}\label{eq:bf}
    \BF_{10} = \frac{p(D\mid \model_1)}{p(D \mid \model_0)} \enspace.
\end{equation} Here, $p(D\mid \model_1)$ and $p(D\mid \model_0)$ are the marginal likelihoods of the two models with their respective parameters integrated out. The Bayes factor indicates how much more likely the data are given the discontinuous model, compared to the continuous model~\citep{Jarosz2014}. Unlike a $p$-value, it can provide evidence for either model, so that it is possible to find evidence supporting the \emph{absence} of a discontinuity~\citep{Wagenmakers2007,Goodman1999a}. Furthermore, this model comparison approach automatically accounts for model complexity~\citep{MacKay2002}.

In the null model, all probability mass of $p(d\mid D, \model_0)$ is concentrated at $d=0$, while for the alternative model we have an effect size distribution $p(d\mid D, \model_1)$. Existing regression discontinuity methods focus on inference of $d$, and hence implicitly condition on $\model_1$. This approach ignores the uncertainty in the model posterior $p(\model\mid D)$, which results in an overconfident overestimate of the effect size, and consequently of too optimistic conclusions of the efficacy of an intervention. This uncertainty can be accounted for via the Bayesian model average (BMA) estimate of $d$:
\begin{equation}\label{eq:bma}
    p(d\mid D) = \sum_{j=0,1} p(d\mid D, \model_j) p(\model_j\mid D) \enspace.
\end{equation} The resulting distribution integrates out the uncertainty of the model, which has been shown to lead to optimal predictive performance~\citep{Hoeting1999}. Since the effect size is by definition zero according to $\model_0$, Eq.~\eqref{eq:bma} is a spike-and-slab distribution that combines a spike at $d=0$ with a Gaussian distribution determined by $\model_1$, where each component is weighted by the posterior probability of the corresponding model. Compared to the overconfident estimation of $d$ conditioned only on $\model_1$, this has a regularizing effect~\citep{Brodersen2015}, shrinking small effect size estimates towards zero. 

\subsection{The continuous model}

The continuous (null) model $\model_0$ implies that the regression does not depend on the threshold, which leaves us with a single regression for all data points. We assume Gaussian observation noise:
\begin{equation*}
    y_i \sim \dnorm{f_0(x_i)}{\sigma_n^2} \enspace.
\end{equation*} Here, $\sigma_n^2$ is the observation noise variance, and $f_0(x_i)$ captures the relationship between the predictor and the response. We do not impose a parametric form on $f_0$, and instead assume $f_0$ follows a Gaussian process (GP)~\citep{Rasmussen2005}:
\begin{equation*}
    f_0 \mid \model_0 \sim \dgp{\mu(x;\theta_0)}{k(x,x';\theta_0)} \enspace,
\end{equation*} with mean function $\mu(x;\theta_0)$ and covariance function $k(x,x';\theta_0))$. We omit the dependence on the hyperparameters $\theta$ when confusion is unlikely to arise.

\subsection{The discontinuous model}

In the alternative model we assume the latent processes before and after $x_0$ are independent. We write
\begin{equation}
    f_1 \mid \model_1 \sim \dgp{\mu(x;\theta_1)}{k_1(x,x';\theta_1} \enspace,
\end{equation} where $k_1(x,x';\theta_1) = k(x,x';\theta_1)$ if $x$ and $x'$ are on the same side of $x_0$, and $k_1(x,x';\theta_1)=0$ otherwise. As a result, the Gram matrix with elements $\vec{K}_{ij}=k_1(x_i, x_j;\theta_1)$ is block-diagonal: 
\begin{equation}
    \vec{K} = \begin{bmatrix} \vec{A} & 0 \\ 0 & \vec{B}
    \end{bmatrix} \enspace,
\end{equation} with the elements in the matrices $\vec{A}$ and $\vec{B}$ corresponding to the covariances between observations at the same side of the threshold $x_0$.

\subsection{Regression discontinuity effect size}

Since $f_1$ is continuous everywhere except at $x_0$, we can determine the effect size given $\model_1$ by taking the difference of its limits as in Eq.~\eqref{eq:rd_effectsize}. The result is a Gaussian distribution:
\begin{equation}\label{eq:es}
    p(d\mid D, \model_1) = \dnorm{m}{s^2} \enspace,
\end{equation} with $m=\lim_{x\downarrow x_0}f_1(x) - \lim_{x\uparrow x_0}f_1(x)$ and
$
        s^2=\lim_{x\downarrow x_0}\mathds{V}[f_1(x)] + \lim_{x\uparrow x_0}\mathds{V}[f_1(x)] = 2\sigma_n^2 \enspace,
$ for stationary covariance functions, where $\sigma_n^2 \in \theta_1$ represents the observation noise hyperparameter of the discontinuous model. 

\subsection{Interrupted time series effect size}

In contrast to RD design, in ITS the discontinuity may induce a nonstationarity in the latent process, such as a change in length-scale or frequency. To address this, we allow the hyperparameters pre- and post-intervention to differ, i.e. $\vec{A}_{ij}=k(x_i, x_j;\theta^{\vec{A}}_1)$ and $\vec{B}_{ij}=k(x_i, x_j;\theta^{\vec{B}}_1)$. The differences in design also imply a different notion of effect size, which is now a function of $x$:
\begin{equation}\label{eq:m1_es}
    p(d(x)\mid D, \model_1) = \dnorm{m(x)}{s_n^2} \enspace,
\end{equation} with $m(x) = f_1\left(x;\theta^{\vec{B}}_1\right) - f_1\left(x;\theta^{\vec{A}}_1\right)$ and $s_n^2=(\sigma_n^{\vec{A}})^2 + (\sigma_n^{\vec{B}})^2$. Note that $s_n^2$ does not depend on $x$.

Of particular interest in ITS are covariance functions that capture long-range correlations, because these have the potential to extrapolate better and hence provide more accurate effect size estimates. The spectral mixture kernel was designed for this purpose~\citep{Wilson2013}. It is defined as a mixture of Gaussian components in the frequency domain:
\begin{equation}
    S(\omega) = \sum_{q=1}^Q w_q \frac{1}{\sigma_q\sqrt{2\pi}} \exp\left[-\frac{1}{2} \left(\frac{\omega-\mu_q}{\sigma_q}\right)^2  \right] \enspace, 
\end{equation} where $\mu_q$ and $\sigma_q^2$ are the mean and variance of each component, respectively. This spectral representation is then transformed into a regular stationary covariance function using the inverse Fourier transform~\citep{Bochner1959}, which results in
\begin{equation}
    k(\tau) = \sum^{Q}_{q=1} w_q \cos\left(2\pi\tau \mu_q \right) \exp\left(-2\pi^2 \tau^2 \sigma^2_q\right) \enspace,
\end{equation} with $\tau=|x-x'|$. The hyperparameters $\theta=(Q, \vec{\mu}, \vec{\sigma}, \vec{w})$ have the following meaning: $Q$ is the number of mixture components, $\mu_q$ indicates the mean frequency of component $q$, the inverse of the variance $1/\sigma_q$ can be interpreted as the length-scale of each component, reflecting how quickly that frequency contribution changes with the input $x$, and the weights $w_q$ determine the relative contribution of each component~\citep{Wilson2013}.

\subsection{Model training}

The marginal likelihood of Gaussian process regression with Gaussian observation noise is available in closed form~\citep{Rasmussen2005}, but unfortunately this is not the case for the model marginal likelihood that integrates over the hyperparameters $\theta$, which is needed to compute the Bayes factor. We therefore approximate these using the Bayesian Information Criterion (BIC)~\citep{Schwarz1978}, given by as
\begin{equation}   
    \log p(D\mid \model_i)\approx \log p(\vec{y}| \vec{x}, \hat{\theta}, \model_i) - \frac{l}{2} \log n \enspace,       
\end{equation} with $\vec{x} = (x_1, \ldots, x_n)^T$ and $\vec{y}=(y_1, \ldots, y_n)^T$,  $l$ the number of hyperparameters, and $\hat{\theta} = \argmax_{\theta}  p(\vec{y} \mid \vec{x}, \theta, \model)$ the optimized hyperparameters, for $i\in\{0,1\}$.

\method is implemented in Python using GPflow 2.2~\citep{GPflow2017}. We set the prior function to the empirical mean. The BMA distribution is approximated via Monte Carlo, and visualized with kernel density estimation. Code and data are available at \href{https://github.com/TODO/CAMERAREADY}{Github}. More details on the training and initialization of the spectral mixture covariance function are provided in Appendix~\ref{app:smc}.

\section{Covariance functions as design choices}

\label{sec:kerneltheory}

The choice of the Gaussian process covariance functions plays two conceptually distinct roles in \method. First, our choice of covariance function reflects our beliefs about the latent process that generated the observations. In traditional RD designs, one assumes a parametrized model such as (local) linear regression. In \method, this explicit parametric form is replaced by a GP prior that assigns a probability distribution to the space of functions. For instance, we may expect functions to be smooth in $x$, or assume functions are a superposition of sine waves~\citep{Rasmussen2005,Wilson2013}. \method can replicate parametrized models by selecting degenerate covariance functions, such as a linear covariance function. 

These modeling choices are crucial in RD design as model misspecification can lead to incorrect inference. When we do not have clear prior beliefs about a covariance function, we may compute the Bayesian model average~\citep{Hoeting1999} across a set of candidate kernels $\mathcal{K}$:
\begin{equation}\label{eq:bftotal}
    \BF_{10}^{\text{total}} = \frac{p(D\mid \model_1)}{p(D \mid \model_0)}=\frac{\sum_{k \in \mathcal{K}} p(D \mid k) p(k \mid \model_1)}{\sum_{k \in \mathcal{K}} p(D \mid k) p(k \mid \model_0)} \enspace.
\end{equation} Here, the quantity $\BF_{10}^{\text{total}}$ serves as a final decision metric to determine an effect in a quasi-experimental design, while a detailed report is provided by inspecting the Bayes factors corresponding to each considered covariance function. Similarly, we can compute a marginal effect size across all considered kernels. In practice, the evidence of one covariance function can dominate all others, in which case the BMA procedure converges to performing the analysis with the best covariance function only.

The second role of the covariance function choice is that it determines to which types of discontinuities \method is sensitive. Importantly, different covariance functions can be used to test fundamentally different hypotheses, as they determine which features of the latent function are part of the alleged effect. For example, the simplest (degenerate) covariance function, the constant function, is sensitive only to differences in the means of the two groups (resulting essentially in a quasi-experimental Bayesian t-test), while the linear covariance function is sensitive to both the difference in mean as well as in slope. In the non-degenerate case, the Mat\'{e}rn covariance function with parameter $\nu=p+1/2$ can detect discontinuities in up to the $p$-th derivative. It has two interesting special cases: one is the exponential covariance function (Mat\'{e}rn with $p=1/2$), which detects only discontinuities in the function itself (and not in its derivatives). This is the nonparametric counterpart of traditional linear regression discontinuity. On the other end is the exponentiated-quadratic covariance function which (Mat\'{e}rn kernel with $\nu=\infty$). This allows us to detect discontinuities of any order, although the amount of data required to detect such subtle effects may become prohibitively large.

\section{Simulations}
\label{sec:sim}

\begin{figure*}[tb]
    \begin{center}
    \centerline{\includegraphics[width=\textwidth]{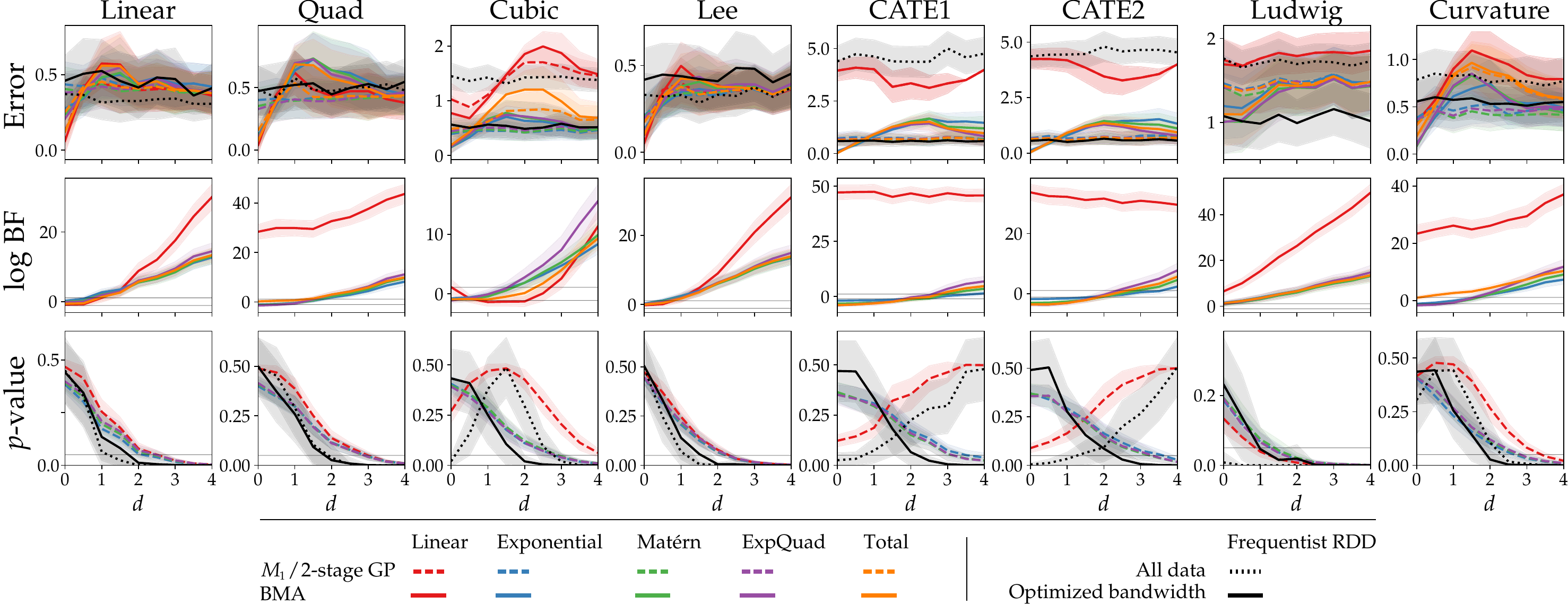}}
    \caption{Simulation results. Top row: the error between the true and estimated effect size. The dashed line indicates the 2-stage GP approach (see text), which is equivalent to $\model_1$. Middle row: The log Bayes factor. Final row: The $p$-values obtained by the RDD baseline (black) and the 2-stage approach. The horizontal dashed lines indicates the common thresholds of $|\BF|<3$ and $p=0.05$.}
    \label{fig:sim_rdd}
    \end{center}
    \vspace*{-5mm} 
\end{figure*}

We evaluate the performance of \method using simulations, using the functions discussed by~\citet{Imbens2012}, which consist of different polynomials up to order 5. We apply \method using linear, exponential, Mat\'{e}rn ($\nu=3/2$) and exponentiated-quadratic covariance functions, as well as the model average of this set. We compare the performance of \method with two baselines. The first is the  \href{https://github.com/evan-magnusson/rdd}{Python RDD package}, which uses linear regression together with the bandwidth selection method by \citet{Imbens2012} to select only a subset of the data around $x_0$ to perform the analysis on. The second is the approach by~\citet{Branson2019}, which first estimates the conditional effect size distribution $p(d\mid \model_1, D)$ and then tests the null hypothesis $d=0$ using this distribution. We refer to this approach as the 2-stage GP. Simulation details and visual examples are provided in Appendix~\ref{app:simulations}. 

Figure~\ref{fig:sim_rdd} shows the absolute difference between the true effect size and the posterior expectations, as well as the decision metrics (averaged over 100 runs). The discontinuous model overestimates $d$ when the true effect size is small, as is to be expected from the implicit conditioning on an effect. The BMA does not have this bias, resulting in lower errors for small and absent effects. For medium effects, this itself can result in a bias due to shrinkage (e.g. the Cubic function), while for large effects the BMA converges to $\model_1$ and the bias disappears. Generally, \method performs on par with the optimized-bandwidth baseline, with worse performance for the Ludwig function, and better for e.g. Curvature, as well as for most cases with an absent or small effect. 

The decision metrics show that for small or absent effects, \method can report evidence in favor of the null, while the corresponding $p$-values are inconclusive. The methods positively identify effects at roughly the same true effect sizes. An interesting special case is observed for the Lee and Ludwig functions, which both feature a discontinuity in their derivative~\citep{Branson2019}, which is correctly picked up by \method even when the magnitude of the effect is small, confirming the ability to detect discontinuities of higher orders.

\begin{figure*}[tb]
    \begin{center}
    \centerline{\includegraphics[width=\textwidth]{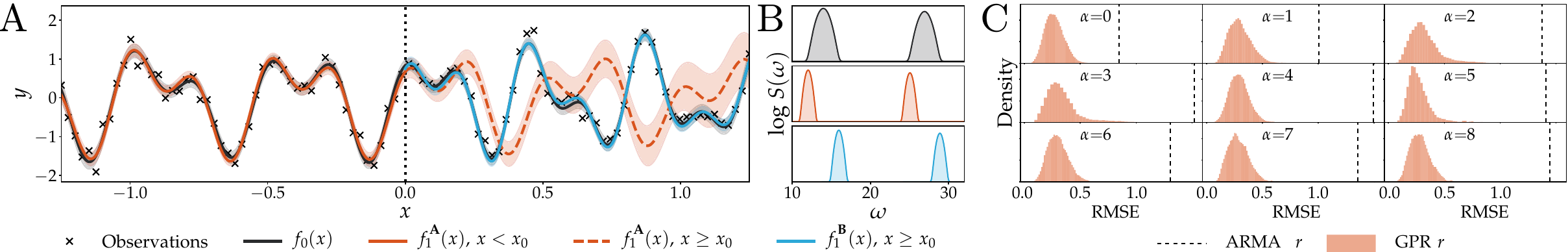}}
    \caption{\textbf{A}. ITS application. Model fit and extrapolation of $\model_0$ and $\model_1$. The data were generated with a post-intervention frequency shift of $\alpha=4$. We find $\log \BF=0.15$. The shaded interval represents two standard deviations around the mean. \textbf{B}. Estimated power spectra. The colors of the power density spectrum correspond to the legend of the regression. \textbf{C}. The RMSE between the estimated and true $d_{\text{ITS}}$ using posterior samples of \method and an ARMA baseline (dashed line).}
    \label{fig:sim_its}
    \end{center}
    \vspace*{-5mm} 
\end{figure*}

We explore the ITS application of \method in another simulation. Here, we generate oscillating data where for $x\geq x_0$ a frequency shift is introduced. We compare our extrapolations based on the spectral mixture covariance function with an ARMA model, which is commonly used in ITS designs~\citep{Prado2010,Jandoc2015}. Details of the simulation procedure are provided in Appendix~\ref{app:simulations}. An example simulation run and \method application is shown in Fig.~\ref{fig:sim_its}A, with a post-intervention frequency shift of $\alpha=4\text{Hz}$. The model correctly recovers the true power spectrum, as well as the decreased amplitude of the second harmonic component post-intervention, and finds barely worth mentioning evidence in favor of an effect ($\log \BF=0.15$). The estimated spectral mixture of the continuous model is centered between the true frequencies of the control and intervention group (Fig.~\ref{fig:sim_its}B). This faithfully represents the null hypothesis that the observations can be explained without any changes in spectral content. As the discontinuity grows larger, the standard deviation of the components of the continuous model increases as well, since it has to account for a larger difference. $\model_1$ instead correctly identifies the true mixture components. Fig.~\ref{fig:sim_its}C shows the RMSE of samples from the posterior distributions of $f_1$ and the true function, as well as the ARMA estimate. \method consistently outperforms the baseline.

\section{Applications}
\label{sec:applications}

\subsection{The effects of winning an election and longevity}

\citet{Barfort2020} investigated the effect of running for US gubernatorial office on longevity. The authors use a regression discontinuity design, and conclude that politicians winning a close election live 5 to 10 years longer than if they had lost. These findings have been heavily criticized~\citep{Gelman2020}, and it is unclear whether a regression discontinuity analysis is actually appropriate here, as there is no clear intervention at $x=0$ (where $x$ is the percentile difference in election result). Despite these concerns, we analyze this data set here as it allows us to demonstrate some of the functionality specific to \method. The data are available from the original publication~\citep{Barfort2020}, and preprocessed following~\citet{Gelman2020}. 

\begin{figure}[tb]
    \begin{center}
    \centerline{\includegraphics[width=0.9\linewidth]{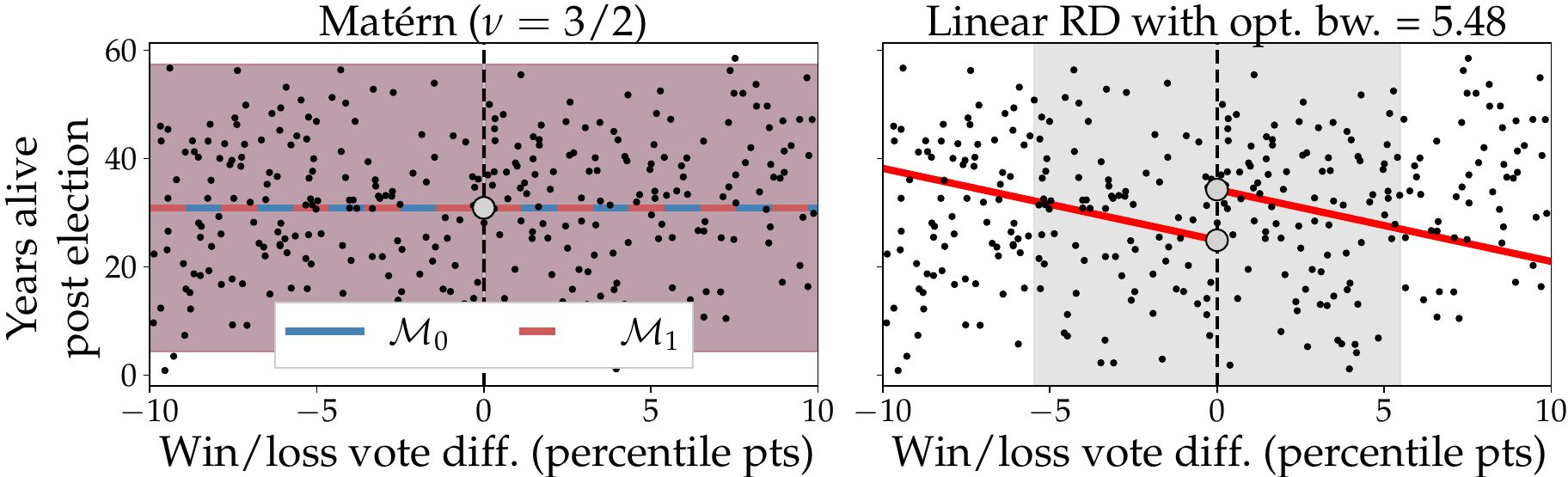}}
    \caption{Discontinuity analysis of the effect of close gubernatorial elections on longevity~\citep{Barfort2020}. Shown are regressions by \method using a Mat\'{e}rn covariance function, and a linear RD baseline with an optimized bandwidth of 5.48 percentile points (shaded area). For \method, the regressions for $\model_0$ and $\model_1$ are nearly identical. For the baseline, the bandwidth optimization leads to a poor linear fit, and hence a spurious detection of an effect.}
    \label{fig:longevity}
    \end{center}
    \vspace*{-5mm} 
\end{figure}

Using the linear RDD baseline we find an optimal bandwidth of 5.48 percentile points using the Imbens-Kalyanaraman procedure~\citep{Imbens2012}. When using this bandwidth and testing for an effect, we find $p=0.019$ and an estimated effect size of 9.4 years. With \method, using either a linear, exponential, or Mat\'{e}rn ($\nu=3/2$) covariance function, we find a more parsimonious explanation of these data by a constant function and a substantial noise term $\sigma^2_n$, as shown by log Bayes factors of -0.12, 0.0, and 0.0, respectively. This indicates that from these data, no clear conclusion can be drawn, and that such a scenario is clearly identified using \method.

\subsection{Phantom border effect on Dutch government elections}

\begin{figure*}[tb]
    \begin{center}
    \centerline{\includegraphics[width=\textwidth]{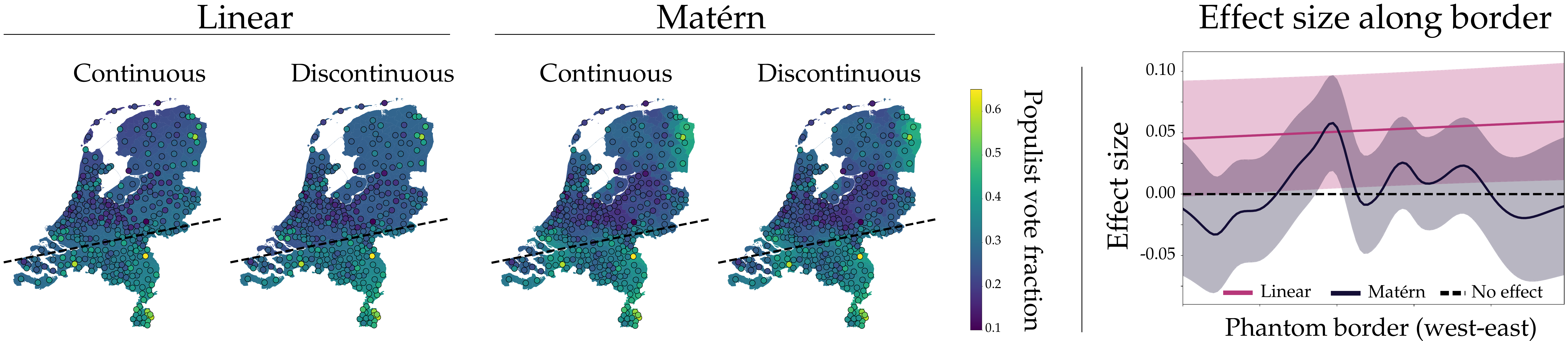}}
    \caption{Discontinuity analysis along a two-dimensional boundary (indicated by the dashed line). \textbf{A}. Circles indicate the observed fraction of populist votes; municipalities are shaded according to the Gaussian process predictions. \textbf{B}. The distribution of effect size conditioned on $\model_1$, $p(d\mid D, \model_1)$, along the phantom border. The shaded interval indicates one standard deviation around the mean.}
    \label{fig:elections}
    \end{center}
\end{figure*}

In 2017, the Dutch general elections were held~\citep{KOOP2018}. According to Dutch electorate geographer De Voogd, the share of votes that go to populist parties\footnote{We refrain from an extensive discussion of the definition of populism and refer to populist parties as those parties that emphasize `an alleged chasm between the elite and the general population'. In the Dutch 2017 elections, parties that fit this description were PVV, SP, 50Plus and FvD~\citep{Mueller2016}.} is different north and south of a so-called `phantom border', a line that historically divided the catholic south of the Netherlands from the protestant north~\citep{DeVoogd2016,DeVoogd2017}. This border serves as a two-dimensional threshold along which one can apply RD design. This special case of RD design where the assignment threshold is a geographical boundary is also referred to as GeoRDD~\citep{Rischard2021}. Here, we test the hypothesis by De Voogd. 

We apply \method using the linear and first-order Mat\'{e}rn covariance functions. The results of the analysis are shown in Fig.~\ref{fig:elections}. The figure shows the Netherlands with the fraction of populist votes per municipality superimposed, together with the phantom border representing the supposed divide in voting behaviour. 

If we assume a linear underlying process, there is strong evidence for a discontinuity ($\log \BF = 24.4$), confirming the hypothesis by De Voogd. Visually however, the data do not appear to follow these linear trends. The nonparametric Mat\'{e}rn covariance function results in evidence \emph{against} an effect ($\log \BF = -3.5$). As the Mat\'{e}rn covariance function fits the data much more accurately than the linear covariance function, the Bayesian model average is completely dominated by the former, leading to the conclusion that the historical phantom border does not create a geographic discontinuity in populist voting behaviour.

\subsection{Kundalini meditation effect on heart rate}

\begin{figure}[tb]
    \begin{center}    
    \vspace*{-5mm} 
    \centerline{\includegraphics[width=\columnwidth]{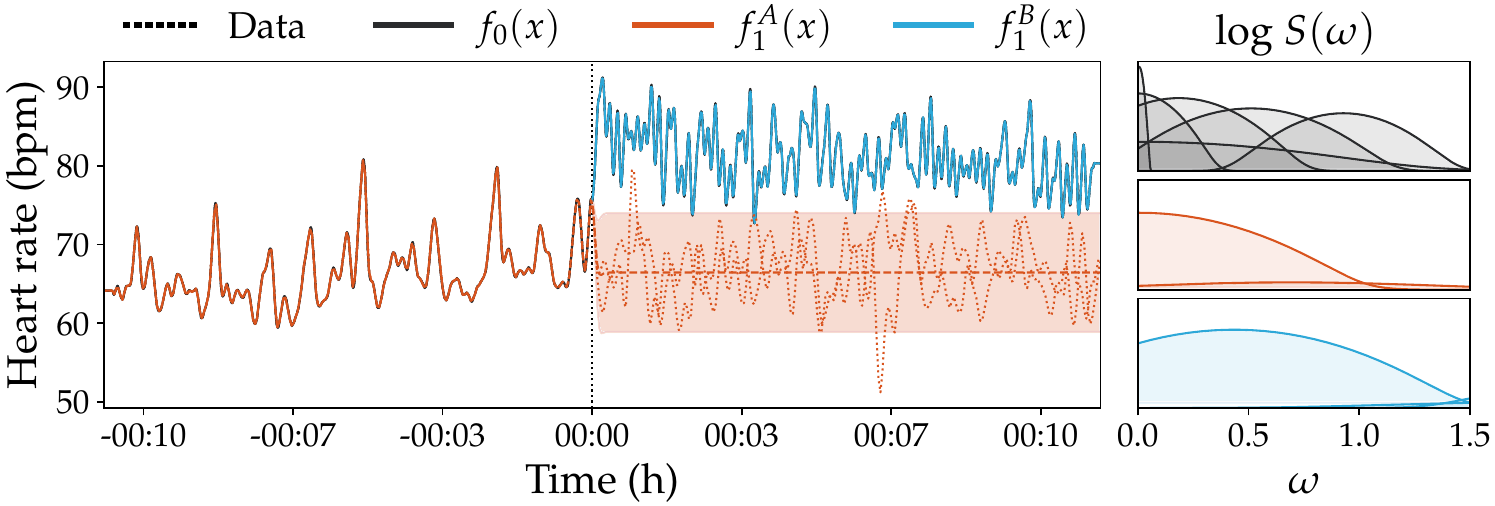}}
    \caption{Analysis of meditation effect on heart rate. Shown is the participant's heart rate, who starts meditation at $x_0=00:00$. The extrapolation, indicated by the dashed (mean) and dotted (posterior samples) red lines, is poor in comparison to the actual observations, which is corroborated by the large log Bayes factor. The panel on the right shows the (log) power spectra expressed by the optimized covariance function hyperparameters.}
    \label{fig:meditation}
    \end{center}
    \vspace*{-5mm} 
\end{figure}

\citet{Peng1999} studied the hypothesis that Kundalini Yoga meditation techniques reduce one's heart rate. However, they find the opposite; the meditation instead brings about an increase in heart rate. The experiment lends itself well for ITS design, but in practice may be difficult to perform because the data are not evenly sampled. However, this is not a prerequisite for Gaussian process regression, which together with the spectral mixture kernel~\citep{Wilson2013} is well-suited to model these data. The observations are obtained from the PhysioNet database and consists of heart rates of two women and two men, of ages 20--52 (mean 33)~\citep{Goldberger2000}. We focus on one participant due to space constraints. Since we do not merely want to detect a change in absolute heart rate, but in its fluctuations, use a changepoint mean function~\citep{Saatci2010} for $\model_0$ and two separate constant mean functions for $\model_1$ to capture the different means. Figure~\ref{fig:meditation} shows the corresponding regression and extrapolation. The continuous model requires more spectral mixture components; $Q=6$ for $f_0$ compared to $Q=2$ for $f\left(x;\theta_1^{\vec{A}}\right)$ and $Q=3$ for $f\left(x;\theta_1^{\vec{B}}\right)$. The analysis finds overwhelming evidence for an effect ($\log \BF = 281.2$).

\section{Discussion} 
\label{sec:discussion}

\method extends naturally to the setting of multiple assignment variables\citep{Reardon2012,Papay2011,Wong2013,Choi2017,Choi2018}. A special case of such multivariate QED is GeoRDD, in which the two-dimensional assignment variable represents a spatial location~\citep{Rischard2021,Keele2015,Keele2017}. Our approach does not assume a univariate threshold to determine the assignment to intervention and control group, and can work on arbitrary complex label functions. This can be a geographical border, but also more complex shapes such as, for example, one region versus the rest of a country, or a particular regime of diagnostic variables.
  
In order to infer causality from QED, one assumes that the alleged change occurs at the threshold, but that the latent process is otherwise stationary. Consequently, the behaviour of the two groups changes sharply around the intervention. In standard RD studies, this locality is controlled via a bandwidth parameter that determines the sensitivity of the detection approach~\citep{Bloom2012}. This requires the availability of sufficient data around the threshold, and the analysis is sensitive to this parameter. In \method with stationary nonparametric covariance functions, the bandwidth is replaced by a length-scale hyperparameter, which we optimize using the model marginal likelihood. The length-scale regulates how fast the correlations between consecutive points decay with their distance, and thus how sensitive \method is around the threshold~\citep{Duvenaud2014}. This implements a trade-off between estimation reliability and the locality assumption that is needed to draw causal inferences~\citep{Geneletti2015}. The benefit of a length-scale instead of a fixed bandwidth parameter is that the relative influence of observations decreases gradually as they are further away from the intervention point, and that this distance is automatically adjusted.

With an exponential covariance function the most rigorous form of locality can be enforced. Here, the Markov properties of the Gaussian process guarantee that only discontinuities at the intervention threshold are detectable. On the other hand, non-local covariance functions such as the periodic covariance function are vulnerable to false positives if the true process is non-stationary. Here, the presence of change points away from the intervention threshold can lead to false alarms, due to the flexibility of the regressions. In this case, or in exploratory applications, \method can be performed in a sliding-window fashion to ensure that the highest Bayes factor is at the intervention threshold. 

\method can be extended in several ways. For instance, we do not currently account for covariates that may serve as confounds for causal inference~\citep{Harris2004,Brodersen2015}. However, such covariates can be explicitly taken into account in the regression models, or even be learned from the observations~\citep{Kocaoglu2017}. Covariate selection can be performed using automatic relevance determination~\citep{Wipf2008}, where we learn separate length-scales for each covariate. Furthermore, improvement is expected from more accurate estimators of the model marginal likelihood than the BIC, such as the ELBO or bridge sampling~\citep{Fourment2020}.

In all, \method serves as a Bayesian nonparametric approach for causal inference in quasi-experimental designs. By selecting the appropriate covariance function, one has precise control over the type of discontinuity that can be detected, as well as a priori assumptions of the latent data generating processes.



\bibliographystyle{plainnat}
\bibliography{library}

\appendix

\section{Training the spectral mixture kernel}
\label{app:smc}

The number of mixture components $Q$ in the spectral mixture kernel of our ITS approach is optimized in the same way as other covariance function parameters are optimized, that is, by optimizing the GP marginal likelihood. The covariance function mixture parameters are initialized by fitting a Gaussian mixture model to the empirical spectral using the Lomb-Scargle periodogram, which is applicable for detecting spectral features in (potentially) unevenly sampled data~\citep{vanderplas2018}. 

\section{Simulations}
\label{app:simulations}

\subsection{RD simulations}

\begin{figure*}[tb]
    \begin{center}
    \centerline{\includegraphics[width=\textwidth]{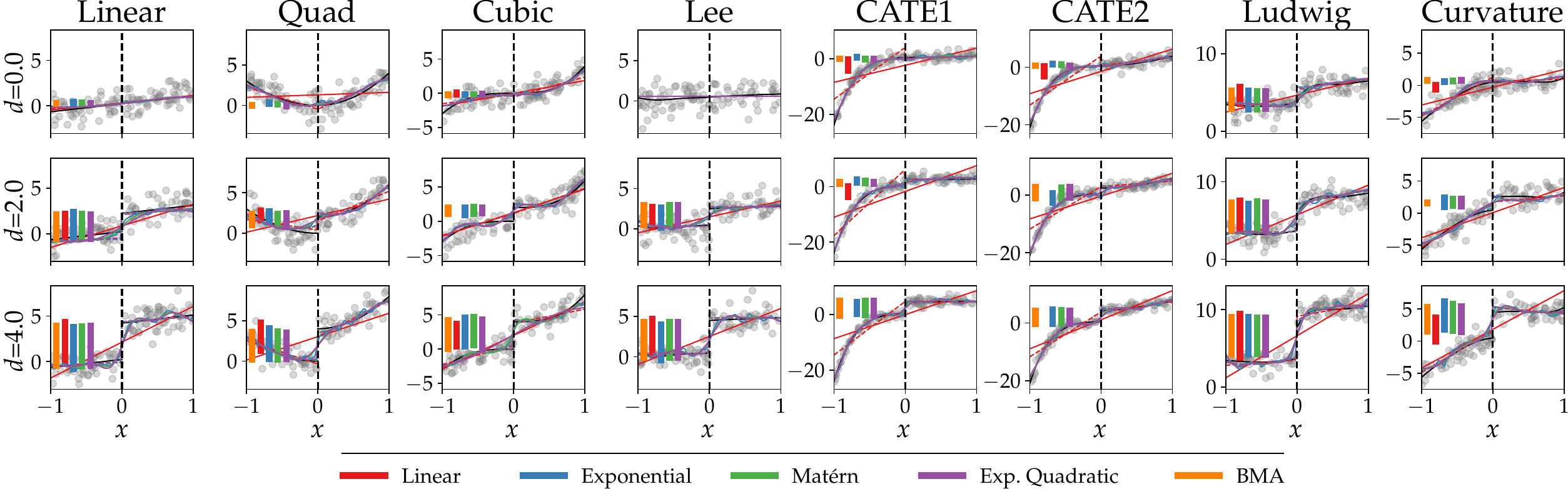}}
    \caption{Example of one simulation run for effect sizes $d\in\{0.25, 1.0, 4.0\}$ and $\sigma=1.0$. The covariance functions used here are linear, exponential, Mat\'{e}rn ($\nu=3/2$) and exponentiated-quadratic. The vertical bars indicate the estimated effect sizes by the discontinuous models for the different covariance functions. As the figure shows, the linear covariance function tends to have the strongest bias, in particular in the low signal-to-noise regime.}
    \label{appfig:examples_sim}
    \end{center}
\end{figure*}

Figure~\ref{appfig:examples_sim} shows an example run of \method on the functions considered in our simulation. Here, the different functions are shown, as well as the regressions by both the continuous and discontinuous models, for each of the four considered covariance functions. The vertical bars in the figure show the expectation of the estimated effect size $p(d\mid D, \model_1)$. These functions used are provided in~\citet{Imbens2012}, and are complemented by a simple linear function to see the behaviour when the linearity assumption by the baseline is actually correct. The function definitions are given by

\begin{flalign*}
     &  & f(x)  & = 0.23+0.89x &  & \text{\llap{Linear}} &\\
     &  & f(x)  &= \begin{cases}
        3x^2 & \text{if $x < x_0$,} \\
        4x^2 & \text{otherwise.}
        \end{cases}  &  &  \text{\llap{Quad}} &\\
    &  & f(x)  & = \begin{cases}
        3x^3 & \text{if $x < x_0$,} \\
        4x^3 & \text{otherwise.}
        \end{cases}  &  & \text{\llap{Cubic}} &\\
    &  & f(x)  & = \begin{cases}
        0.48 + 1.27x + 7.18x^2 + 20.21x^3 + 21.54x^4 + 7.33x^5 & \text{if $x < x_0$,} \\
        0.48 + 0.84x - 3.0x^2 + 7.99x^3 - 9.01x^4 + 3.56x^5 & \text{otherwise.}
        \end{cases}   &  & \text{\llap{Lee}} &\\
    &  & f(x)  & = 0.42 + 0.84x - 3.0x^2 + 7.99x^3 - 9.01x^4 + 3.56x^5 &  & \text{\llap{CATE1}} &\\
    &  & f(x)  & = 0.42 + 0.84x + 7.99x^3 - 9.01x^4 + 3.56x^5 &  & \text{\llap{CATE2}} &\\
    &  & f(x)  & = \begin{cases}
        3.71 + 2.3x + 3.28x^2 + 1.45x^3 + 0.23x^4 + 0.03x^5& \text{if $x < x_0$,} \\
        3.71 + 18.49x - 54.81x^2 + 74.3x^3 - 45.02x^4 + 9.83x^5& \text{otherwise.}
        \end{cases}  &  & \text{\llap{Ludwig}} &\\
    &  & f(x)  & = \begin{cases}
        0.48 + 1.27x - 3.44x^2 + 14.147x^3 + 23.694x^4 + 10.995x^5  & \text{if $x < x_0$,} \\
        0.48 + 0.84x - 0.3x^2 - 2.397x^3 - 0,901x^4 + 3.56x^5  & \text{otherwise.}
        \end{cases}  &  & \text{\llap{Curvature}} &\\
\end{flalign*}

For each latent function $f$, we generate $n=100$ observations $(x_i, y_i)$ according to the following procedure:
\begin{equation*}
    \begin{split}
        x_i &\sim \mathcal{U}(-1, 1) \\
        y_i \mid x_i, \sigma, d, f &\sim \dnorm{f(x_i) + d[x_i\geq x_0]}{\sigma^2} \enspace,
    \end{split}
\end{equation*} where the threshold $x_0=0$. We fix $\sigma=1.0$ and vary $d\in \{0, 0.5, \ldots, 4.0\}$, effectively providing a range of different signal-to-noise regimes. Next, we subject the simulated data to analysis by \method, using a first-order polynomial, an exponential, a Mat\'{e}rn ($\nu=3/2$) and a exponentiated-quadratic covariance function. For each covariance function, we compute the Bayes factor for the presence of a discontinuity, and we estimate the marginal effect size (Eq.~(7) in the main text). 

We also compute $p$-values and conditional effect size estimates for the RDD baseline using the Imbens-Kalyanaraman bandwidth optimization method, and for the two-step GP method~\citep{Branson2019}. The latter corresponds to fitting a GP regression pre- and post-intervention, computing the implied effect size distribution at $x=x_0$, and then testing whether $d=0$ is in this distribution. Note that this distribution is the same as our conditional effect size distribution in Eq.~(4), but the subsequent testing procedure is different; a frequentist significance test is performed rather than Bayesian model comparison. 

For each of the three methods, we compute the absolute error between the estimated and true effect size, and use this to quantify the estimation performance. In addition, we show the decision metrics for each method, so that the differences between a Bayesian and frequentist approach can be seen.

As these and the aggregated results (see main text) show, the linear covariance function results in the largest bias in the effect size estimate, which is unsurprising given its strong assumptions that do not match the data generating process here. 

\subsection{ITS simulations}

The latent function for the ITS simulation is given by
\begin{equation*}
    f(x) = \begin{cases}  \sin(12x) + \tfrac{2}{3}\cos(25x) & \text{for $x < x_0$ and} \\
    \sin((12+\alpha)x) + \tfrac{2}{3}\cos((25+\alpha)x) & \text{for $x\geq x_0$,}
    \end{cases}
\end{equation*} with $x_0=0$, and where $\alpha$ indicates the shift in frequency (set to $\alpha=4$ in the example figure in the main text). We vary $\alpha$ across the range $[0, \ldots, 8]$ Hz. For observation noise, we once more assume
\begin{equation*}
    y \sim \mathcal{N}(f(x), \sigma^2) \enspace,
\end{equation*} and $\sigma^2=0.2$. For each value of $\alpha$, we generate 20 datasets containing $n=200$ evenly spaced observations. 

As a baseline for comparison of our approach we use an ARMA model~\citep{Prado2010}. The parameters of the ARMA model are determined using a grid search and its BIC score. We then compare the root-mean-squared-error between samples from the predictive distribution obtained by BNDD and the true post-intervention signal, and similarly evaluate the performance of the ARMA extrapolations and the true signal. The results are shown in Fig.~2 in the main text, and demonstrate that the spectral mixture kernel-based ITS approach consistently outperforms the baseline.

\end{document}